\documentclass[conference]{IEEEtran}
\IEEEoverridecommandlockouts

\usepackage{cite}
\usepackage{amsmath,amssymb,amsfonts}
\usepackage[ruled,linesnumbered]{algorithm2e}
\usepackage{graphicx}
\usepackage{textcomp}
\usepackage{xcolor}
\def\BibTeX{{\rm B\kern-.05em{\sc i\kern-.025em b}\kern-.08em
    T\kern-.1667em\lower.7ex\hbox{E}\kern-.125emX}}
\begin{document}

\title{Joint Optimization of Routing and Purification to Meet Fidelity Targets in Quantum Networks}

\author{\IEEEauthorblockN{Gongyu Ni\IEEEauthorrefmark{1}, 
Holger Claussen\IEEEauthorrefmark{1}\IEEEauthorrefmark{2}\IEEEauthorrefmark{3}, Lester Ho\IEEEauthorrefmark{1},}
\IEEEauthorblockA{\IEEEauthorrefmark{1}Tyndall National Institute, Dublin, Ireland}
\IEEEauthorblockA{\IEEEauthorrefmark{2}University College Cork, Ireland}
\IEEEauthorblockA{\IEEEauthorrefmark{3}Trinity College Dublin, Ireland}
Emails: 
\{gongyu.ni, lester.ho, holger.claussen\}@tyndall.ie\vspace{-2em} 
}

\maketitle

\begin{abstract}

Quantum networks rely on high-fidelity entanglement links, but achieving target fidelity often increases latency and Bell pair consumption due to purification. This paper proposes a cost-based scheduler that jointly optimizes path selection and purification round, along with two hop-level estimators (a Deep Neural Network classifier and a Bayesian optimizer) to predict the minimal purification rounds needed for target hop fidelity. The scheme flexibly adjusts final entanglement fidelity while minimizing latency, improving request success rates and efficient Bell pair usage. Simulations integrating purification, entanglement generation, and network-level scheduling show that our approach reduces mean latency by up to 8\% and increases success rates by 14\% compared to fixed-round purification with FIFO scheduling.

\end{abstract}

\begin{IEEEkeywords}
entanglement, fidelity, scheduling, purification, latency, Deep Neural Networks, Bayesian optimization.
\end{IEEEkeywords}

\section{Introduction}


Quantum networks integrate the generation, transmission, and utilization of quantum information. A crucial step is establishing entanglement between communicating nodes. However, transmission errors increase with link length \cite{dur1999quantum}. To mitigate this challenge, quantum repeaters are employed to divide the total distance into shorter segments. Moreover, by consuming more entangled pairs for purification, the lower fidelity entangled pairs convert to higher fidelity pairs.

Although various purification protocols \cite{deutsch1996quantum,bennett1996purification,krastanov2019optimized} have been proposed, purification management at the network layer is still lacking. In \cite{davies2024entanglement}, a buffering system incorporating quantum memory and purification was introduced, showing that consistent purification increases the average fidelity, despite some entanglements being discarded due to purification failures. For two-node systems, \cite{elsayed2024trade} analysed the trade-off between fidelity and request waiting time under a First-In, First-Out queue.

In contrast to link-level entanglement, this paper focuses on the study of purification at the network level. In \cite{victora2023entanglement}, purification was integrated into entanglement routing across different topologies. However, optimizing guaranteed fidelity and latency through joint path selection and entanglement request scheduling remains an open problem.

Different quantum applications require distinct fidelity levels. For instance, in quantum computing, the fidelity of remotely entangled network qubits around 96.89\% is observed \cite{main2025distributed}, while the fidelity threshold for quantum key distribution (QKD) depends on the protocol (ie., device-independent QKD requires a minimum fidelity of 89.2 \% \cite{zhang2022device}). Therefore, it is crucial to develop methods for tuning entanglement fidelity levels with respect to different quantum applications, to achieve the required fidelity while avoiding overheads of providing higher than necessary fidelity. In particular, a more user-centric approach using performance metrics should be considered. 

In this paper, a technique for optimizing purification and entanglement request scheduling based on purification-round estimation to achieve a target fidelity is proposed. Two main contributions are presented: (1) hop-level purification-round estimators (DNN and Bayesian) tied to target fidelity, and (2) a cost-based scheduler that leverages these estimators to jointly optimize path selection and purification. 

\section{System Model}

This section presents a model that integrates quantum protocols and network simulation. 
Entanglement is generated after the routing decision, tightly coupling link and network layers, and jointly considering routing, entanglement, and purification.

\subsection{Purification}


We assume a two-qubit state $\rho$ is in the Werner state \cite{werner1989quantum}:  
\begin{equation}
\rho = p |\phi\rangle\langle\phi| + \frac{1-p}{4} I_4,
\end{equation}  
which consists of a maximally entangled state \( |\phi\rangle \) and a maximally mixed state \( I_4 \), where \( p \) is a mixing parameter, with \( p \in [0,1] \). When $p = 1$, the state is purely entangled. When $p = 0$, the state is fully decoherent.

For purification, it is assumed that multiple Bell pairs with an initial fidelity above $0.5$ because if the initial fidelity is below this lowest initial fidelity value, Werner states can no longer be used for purification \cite{bennett1996purification}.  
 
Measurement is performed on one of the Bell pairs from every two Bell pairs, and the results are transmitted over a classical channel. The Bell pair is kept only if the measurement results match, otherwise, it is discarded.


In each hop, under the assumption of the same initial fidelity \(f\) of these generated Bell pairs, the purification fidelity, \(f_p\) , is given by \cite{munro2015inside}:  
\begin{equation}
\label{eqn:hop_fidelity}
f_p= \frac{f^2 + \frac{1}{9} (1-f)^2}{f^2 + \frac{2}{3} f(1-f) + \frac{5}{9} (1-f)^2},
\end{equation}  
with a success probability of purification, \(s_p\) given by:  
\begin{equation}
s_p = f^2 + \frac{2}{3} f(1-f) + \frac{5}{9} (1-f)^2.
\label{eqn:success_probability}
\end{equation}


\subsection{Entanglement distribution}

Entanglement distribution between two end nodes can occur in two ways: either the nodes are directly connected and perform entanglement purification, or quantum repeaters divide the total distance into shorter segments, each purified before entanglement swapping connects them.


Under the assumption of a chain of $n$ neighbouring Werner states across $n$ connected hops, the output state fidelity $F_c$ is given by the following equation \cite{dur1999quantum}:

\begin{equation}
\label{eqn:chain_fidelity}
F_c = \frac{1}{4} + \frac{3}{4} \left[\frac{p_2\left(4 \eta^2 - 1\right)}{3}\right]^{n-1} \prod_{h=1}^{n} \frac{4 f_{p_h} - 1}{3},
\end{equation}
where $p_2$ is the two-qubit gate fidelity, $\eta$ is the measurement fidelity, and $f_{p_h}$ is the purified fidelity of each hop $h$.

\subsection{Purification round estimation}\label{subsec:estimation_purification_round}


In a node chain, the final fidelity $F_c$, as shown in Eq. (\ref{eqn:chain_fidelity}), between the end nodes is bounded by the minimum fidelity of all purified hops along the path:

\begin{equation}
F_c \leqslant \min \left\{f_{l_0,q_0}, f_{l_1,q_1}, \dots, f_{l_h, q_h}\right\}
\end{equation}
where $f_{l_h,q_h}$ represents the fidelity of the purified hop $h$ between neighbouring nodes $v_{l_h}$ and $v_{q_h}$.

According to Eq. (\ref{eqn:chain_fidelity}), given the final fidelity threshold $F_\theta$ of the two end nodes and the number of hops $n$, the target hop fidelity, denoted as $f_{ph}$, can be estimated.

This paper proposes two methods for estimating the target hop fidelity. The first method (optimistic estimation) focuses on determining the minimum required target hop fidelity under the constraints imposed by other hops with highest hop fidelities $f_{ph_\text{max}}$. 



The second method (conservative estimation) adopts a conservative approach by considering the worst-case scenario, where it is assumed that all hop fidelities have the same target value. This assumption simplifies the analysis by setting all $f_{ph}$ values equal in Eq. (\ref{eqn:chain_fidelity}), and solving for a common $f_{ph}$. This assumption is conservative as it does not include the scenarios in which certain hops are capable of generating high-fidelity Bell pairs, potentially exceeding the target fidelity.


\subsection{Network model} \label{network_model}

For the network simulation, a random topology is generated using the Watts–Strogatz model \cite{watts1998collective}, which captures high local clustering and short average path lengths typical of communication networks. The Watts–Strogatz model provides a controlled way to vary the presence of shortcut edges, while randomised edge lengths capture heterogeneity in physical deployment. Varying the rewiring probability therefore isolates the impact of shortcut density on routing/scheduling.
Time is divided into fixed intervals $I_t$ (time slots). Within each time slot, multiple new entanglement requests arrive, and the number of requests follows a Poisson distribution.


However, only one request can be handled at a time, while the remaining requests are queued. At the end of a time slot, any unserved entanglement requests are carried over to the next time slot and queued ahead of the newly generated requests in the subsequent time slot.

\section{Performance metrics} \label{prob_formulation}



In a quantum network, entanglement requests are queued before scheduling, requiring optimal path selection and scheduling to minimize individual entanglement request latency and maximize the overall successful rate of these requests. The utilization of Bell pairs should be optimized while ensuring that the final fidelity surpasses the target fidelity threshold. Performance is evaluated by four metrics: request latency, fidelity, per-slot success rate, and Bell pair utilization.

\subsection{Latency}

To evaluate the entanglements exceeding the target fidelity, their latencies $l$ are computed using Eq.~\ref{eqn:request_delay_time}.

\begin{equation}  
\label{eqn:request_delay_time}  
    l = t_{f} - t_{g},  
\end{equation}  
where $t_{f}$ denotes the time of request fulfilment, and $t_{g}$ is the time of request generation.  
    
\subsection{Fidelity}

For each entanglement request, once the entanglement between the two end nodes is established, the final fidelity is calculated by Eq. (\ref{eqn:chain_fidelity})  to evaluate the quality of the entanglement. 


\subsection{Success rate}

The success rate $R_s$ is defined as the number of successfully established entanglement requests $n_s$ per time slot. The higher the success rate, the more entanglement requests satisfy the target fidelity requirement.



\subsection{Bell pair utilization rate}

The Bell pair utilization rate $U$ is calculated as the ratio of the total number of low-fidelity Bell pairs $n_b$ consumed by the purification process to the total number of successfully established entanglement requests $n_s$, which characterizes how many Bell pairs are used for every successful request in Eq. (\ref{eqn:utilization}). Given the same number of successful requests, a lower Bell pair utilization rate reflects greater efficiency in Bell pair utilization.

\begin{equation}
\label{eqn:utilization}
U = \frac{n_{b}}{n_{s}}.
\end{equation}


\section{Purification round estimation techniques} \label{sec:round_estimation}


In this section, both the Deep Neural Network (DNN) and Bayesian methods for estimating purification rounds are described. These methods capture the relationship between a hop's required purification rounds and its resulting fidelity. These correlations are then used by the cost-based schedulers. 


\subsection{Training Dataset}



The training dataset for both Deep Neural Network (DNN) and Bayesian estimators, includes each hop's purification round and the resulting fidelity. 

\subsection{Deep Neural Network (DNN) Estimator}

\begin{figure}[ht]
    \centering
    \includegraphics[width=0.8\linewidth]{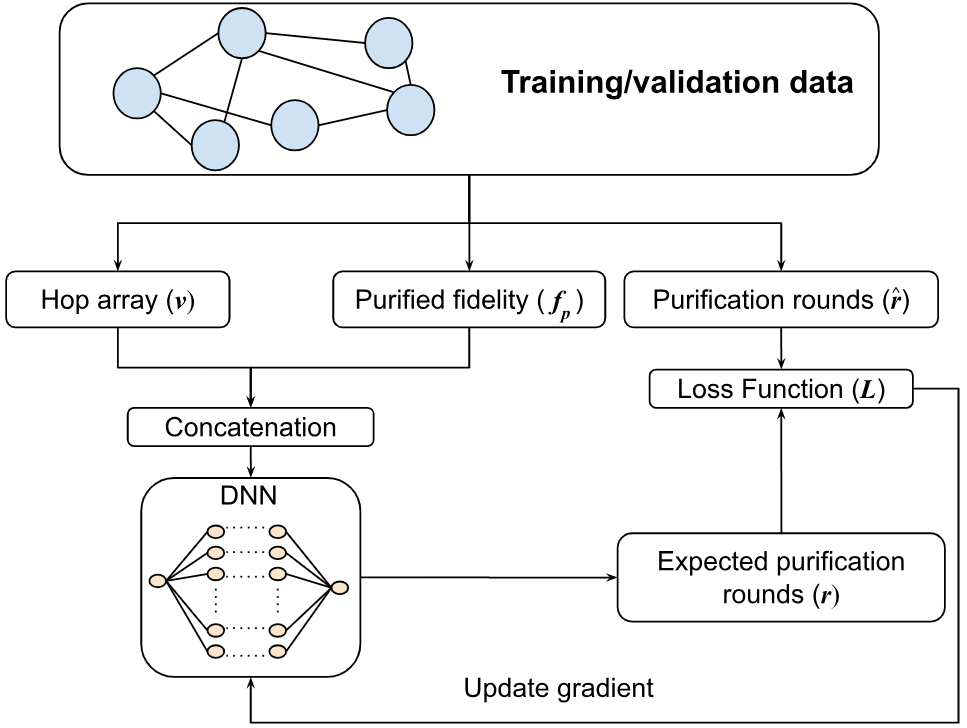}
    \vspace{-3mm}
    \caption{DNN training structure. The dataset consists of $10,000$ data points per hop, splitting into training and validation sets with an $80/20$ ratio.}
    \label{fig:dnn_structure}
\end{figure}

The structure of the DNN model is shown in Fig. \ref{fig:dnn_structure}. 

\subsubsection{Input Space} 


The hop array $v = \left[v_s, v_d\right]$ consists of two binary vectors. Each node $v_m$, which may serve either as the source $v_s$ or destination node $v_d$, is encoded as a binary vector:

\begin{equation}
\label{eqn:single_request}
v_m = 
\begin{cases}
1, & \text{if the current node is } m, \\ 
0, & \text{otherwise}.
\end{cases}
\end{equation}

The purified fidelity $f_p$ is concatenated along with the hop array as input for the training model.

\subsubsection{Output Space} 

The DNN model outputs the predicted number of purification rounds.

\subsubsection{Loss} 

By comparing the model's predicted number of purification rounds with the actual number of purification rounds, the cross-entropy loss is used to train the model.




\subsubsection{Training Result}
Once the model is trained, it is used to predict the required number of purification rounds to perform for a given hop based on the target fidelity. 



\subsection{Bayesian Optimization of purification rounds}


The objective in Bayesian optimization is to optimize the purification rounds required for each hop in order to achieve a target fidelity

With the target fidelity, the Bayesian optimization process restricts the search space to purification rounds $r \in [1,3]$. For each candidate configuration, the mean fidelity $\hat{F}(r)$ is computed from the available dataset corresponding to the given hop and purification round. To measure deviation from the desired fidelity while discouraging excessive resource use, the following penalty function is introduced:

\begin{equation}
\label{eq:penalty}
\mathcal{L}(r) = \left| \hat{F}(r) - F_{\text{target}} \right|,
\end{equation}

where $\hat{F}(r)$ denotes the empirical fidelity under purification round $r$, and $F_{\text{target}}$ is the prescribed fidelity. In cases where no measurement exists for the chosen round, a fixed penalty value is assigned to prevent invalid solutions from being selected.  

For each hop, an independent optimization task is performed using Gaussian-process Bayesian optimization, implemented via \texttt{gp\_minimize} from the \texttt{scikit-optimize} Python library. The optimizer adaptively samples the search space over $30$ evaluations per target fidelity, balancing exploration and exploitation to minimize the penalty function. 



\section{Scheduling methods}

\subsection{Cost scheduling with purification estimation}

Using purification round estimators, paths with the lowest combined sum of path length and total purification rounds are prioritized to minimize entanglement distribution time. The cost function for ranking entanglement requests is:

\begin{equation}
\label{eqn:cost_function}
\begin{aligned}
C &= \gamma D + (1-\gamma) R, \\
\end{aligned}
\end{equation}

where $C$ denotes the request cost. $D$ and $R$ represent the normalized total path length and normalized total purification rounds, respectively. The path length is normalized by subtracting the minimum path length in the graph and dividing by the range between the maximum and minimum path lengths. Similarly, the purification rounds are normalized by subtracting the minimum purification rounds and dividing by the range between the maximum and minimum purification rounds along the path between the two nodes in the graph.

A tunable coefficient \(\gamma \in [0,1]\) balances the influence of path length and purification rounds; its optimal value depends on entanglement, purification times, and network topology.  

To find the lowest-cost path for a source-destination pair, multiple candidate paths are considered, and the path with minimal cost is selected. This procedure is repeated for all source-destination pairs. Requests are then processed sequentially in order of increasing cost until all are served or the time slot expires.

\subsection{Shortest-path fixed purification}



To benchmark the proposed cost scheduling, we consider a shortest-path method with fixed purification. Without purification round estimation, this scheme applies a predetermined number of purification rounds to each hop and prioritizes requests with the shortest path length. The number of purification rounds is fixed to either one or two and is not adaptive. 


\subsection{FIFO}


In the FIFO scheme, requests are processed sequentially in order of arrival. Each request performs one purification round per hop before establishing entanglement.




\section{Performance evaluations}


This section describes the simulation settings and evaluates the proposed scheduling methods against the shortest-path and FIFO baselines using the latency, fidelity, success rate, and Bell pair utilization metrics defined in Section~\ref{prob_formulation}.

\subsection{Simulation settings}


For simplicity, all Bell pairs generated within a hop are assumed to have the same initial fidelity \( f_h \), but this fidelity varies across hops. The weight of a hop is denoted by \( w_h \), and its initial fidelity \( f_h \) is randomly selected from a Gaussian distribution $\mathcal{N}$ with a mean of \( 1 - w_h \) and a standard deviation of \( 0.01 \):
\begin{equation}
\label{eqn:generated_fidelity}
f_h \sim \mathcal{N}(1 - w_h, 0.01).
\end{equation}

Each hop distance is uniformly set to $1$ km. The purification time per round, $t_p$, is assumed to be $10 \mu s$, as classical information requires less than $100 \mu s$ to travel $10$ km through optical fibre \cite{davies2024entanglement}.

The estimated time to establish entanglement between two end nodes is calculated based on the number of nodes $n_v$ and the average hop distance $\overline{d}$:

\begin{equation}
t_r = (n_v - 1) \cdot \overline{d} \cdot t_c,
\end{equation}
where $t_r$ represents the estimated runtime to establish an entanglement between the end nodes, and $t_c$ is the estimated runtime for building entanglement per kilometre, which should be longer than the time required for classical information transmission and is assumed to be $10 \mu s$. The entanglement generation time $t_g$ is assumed to be $50~\mu s$, based on a quantum source repetition rate of $R = 10$ MHz \cite{sangouard2011quantum}.



Since the hop distance considered is short, and the durations $t_p$, $t_c$, and $t_g$ are relatively short in comparison with the coherence times reported for entangled memories \cite{bar2013solid,pompili2022experimental}, all successfully generated entanglement pairs are assumed to remain coherent within the considered time slot intervals $I_t$. 

In addition, $\gamma$ in the Eq. (\ref{eqn:cost_function}) is set to $0.3$, which is the optimal value observed through experimentation of the network configuration including $t_c$, $t_p$ and the topology. 

The target fidelity threshold $F_\theta$ is set to $0.83$. This is based on the theoretical threshold for Quantum Key Distribution (QKD) using the Clauser–Horne–Shimony–Holt (CHSH) Bell inequality, which is $2(\sqrt{2} - 1) \approx 0.8284$ \cite{sajeed2020bright}. The mentioned simulation parameters are summarized in Table \ref{tab:par}.


\begin{table}[htbp]
    \centering
    \addtolength{\tabcolsep}{-2.5pt}
    \caption{Simulation Parameters.} 
    \vspace{-0.25cm}
        \renewcommand{\arraystretch}{1.05}
    \label{tab:par}
    \begin{tabular}{c c c}
        \hline
            \hline
        \textbf{Parameter}&\textbf{Description}&\textbf{Value}\\
        \hline
            $p_2$ &Two-qubit gate fidelity &$0.98$\\
            $\eta$ &Measurement fidelity & $0.99$ \\
            $d$ &Hop distance & $1$ km\\
            $n_v$ &Number of nodes &$10$\\
            $w_{h}$ &Hop weight selection list & $[0.05, 0.1,0.15,0.2,0.25]$ \\
            $F_\theta$ &Fidelity threshold& $0.83$ \\
            $N_{t}$	&Number of time slots & $1,000$\\
            $t_{p}$ &Purification time per round & $10 \mu s$ \\
            $t_{c}$ &Entanglement time per Km & $10 \mu s$ \\
            $t_{g}$ &Entanglement generation time & $50 \mu s$ \\
            $I_{t}$ &Time slots intervals & $1000 \mu s$ \\
            $\gamma$ &Tuning Coefficient  & $0.3$\\
            $\lambda$ &Network load selection list & $[2, 6, 8]$ \\	
        \hline
            \hline
    \end{tabular}
\end{table}

\subsection{Results}

The benefits of the proposed cost scheduling with purification round estimation techniques lie in its flexibility to tune fidelity levels, latency management, and intelligent utilization of Bell pairs. The observations that can be concluded from these results are outlined below.

\subsubsection{Flexibility in adjusting fidelity levels}



\begin{figure}
    \centering
    \includegraphics[width=0.45\textwidth]{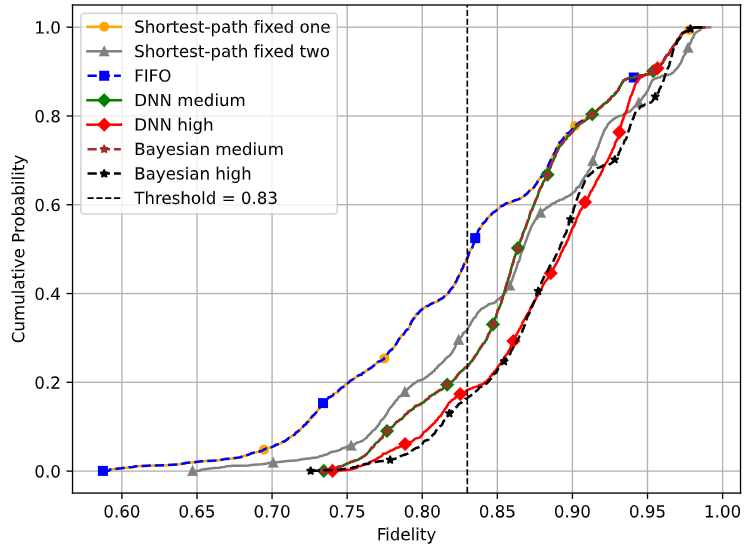}
    \vspace{-6mm}
    \caption{CDF of fidelities across different purification schemes in network load $\lambda = 2$.}
    \label{fig:fidelity}
\end{figure}

Figure~\ref{fig:fidelity} shows the cumulative distribution function (CDF) of the final entanglement fidelities from the simulations. Lower intersection points of the threshold line with the fidelity curves indicate a higher probability of the entanglement exceeding the fidelity threshold. The distributions are largely similar across network loads, making the case with $\lambda = 2$ representative.

The shortest-path fixed purification schemes with one and two rounds are denoted as ``Shortest-path fixed one'' and ``Shortest-path fixed two,'' respectively. Since both the ``Shortest-path fixed one'' and ``FIFO'' strategies use a single purification round per hop, they achieve identical fidelities.

The ``DNN medium'' and ``Bayesian medium'' strategies use optimistic estimates, whereas ``DNN high'' and ``Bayesian high'' employ conservative estimates, as described in Section \ref{subsec:estimation_purification_round}. Since the conservative estimate yields higher target fidelities than the optimistic one, ``DNN high'' and ``Bayesian high'' achieve significantly higher fidelities than their medium counterparts, which shows the cost-scheduling method's ability to adjust fidelity.  


\subsubsection{Intelligence in Bell pair utilization}

Table \ref{tab:utilization_ratio} shows the resulting  Bell pair utilization rate $U$, defined in Eq. (\ref{eqn:utilization}). 
\begin{table}
    \caption{Utilization of Bell pairs per successful request.}
    \vspace{-0.25cm}
    \renewcommand{\arraystretch}{1.05}
    \addtolength{\tabcolsep}{6pt}
    \label{tab:utilization_ratio}
    \centering 
    \begin{tabular}{c c c c}
        \hline
        \hline
        \textbf{Network load ($\lambda$)} & $2$ & $6$ & $8$ \\
        \hline
        \textbf{Shortest-path fixed one} & $32.14$ & $32.44$ & $32.22$ \\
        \textbf{Shortest-path fixed two} & $73.66$ & $75.37$ & $75.06$ \\
        \textbf{FIFO} & $32.14$ & $32.44$ & $32.20$ \\
        \textbf{DNN medium} & $41.99$ & $42.61$ & $42.55$ \\
        \textbf{DNN high} & $86.00$ & $85.16$ & $86.52$ \\
        \textbf{Bayesian medium} & $42.02$ & $42.58$ & $42.59$ \\
        \textbf{Bayesian high} & $78.88$ & $78.53$ & $79.72$ \\
        \hline
        \hline
    \end{tabular}
    \vspace{-0.3cm}
\end{table}
The ``Shortest-path fixed one'' and FIFO methods has the lowest utilization per request, as they always use a single purification round. However, this comes at the cost of lower number of successful entanglement requests in Fig.~\ref{fig:fidelity}. In contrast, both ``DNN high” and ``Bayesian high” consume the largest number of Bell pairs per request for purification but achieve the highest success rate in establishing entanglement between two nodes.

While previous methods favor either low Bell pair usage or high success rates, tuning the purification rounds based on estimation balances both objectives. The ``DNN medium'' and ``Bayesian medium'' approaches handle this trade-off effectively, achieving higher success rates in Fig.~\ref{fig:fidelity} while consuming significantly fewer Bell pairs per successful request compared to the static ``Shortest-path fixed two'' method. This demonstrates more efficient Bell pair utilization through informed estimation of purification rounds.

\subsubsection{Effectiveness in latency optimization}

Methods targeting high fidelities ("DNN high" and "Bayesian high") incur higher latencies in Table \ref{tab:latency_per_success} due to more purification rounds, whereas methods using a single purification round ("Shortest-path fixed one" and FIFO) achieve the lowest latencies. This reflects the latency–fidelity trade-off illustrated in Tables \ref{tab:success_per_timeslot}, where more purification rounds also lead to higher success rates.

In contrast, the DNN medium and Bayesian medium methods balance latency and fidelity, achieving lower latencies than the benchmark ``Shortest-path fixed two'' method while maintaining higher success rates across different network loads (Table~\ref{tab:gains}). These gains are computed as the relative difference in mean values compared to the benchmark.


\begin{table}
    \caption{Mean latency ($\mu s$) per successful request.}
    \vspace{-0.25cm}
    \renewcommand{\arraystretch}{1.05}
    \addtolength{\tabcolsep}{6pt}
    \label{tab:latency_per_success}
    \centering
    \begin{tabular}{c c c c}
        \hline
        \hline
        \textbf{Network load($\lambda$)} & $2$ & $6$ & $8$ \\
        \hline
        Shortest-path fixed one & 155.09 & 297.38 & 376.18 \\
        Shortest-path fixed two & 175.79 & 339.69 & 450.33 \\
        FIFO & 158.86 & 307.64 & 389.08 \\
        DNN medium & 165.85 & 318.28 & 413.25 \\
        DNN high & 183.47 & 353.76 & 493.07 \\
        Bayesian medium & 165.85 & 318.28 & 413.25 \\
        Bayesian high & 185.38 & 358.68 & 500.44 \\
        \hline
        \hline
    \end{tabular}
    \vspace{-0.3cm}
\end{table}

\begin{table}
    \caption{Mean number of successful requests per time slot.}
    \vspace{-0.25cm}
    \renewcommand{\arraystretch}{1.05}
    \addtolength{\tabcolsep}{10.5pt}
    \label{tab:success_per_timeslot}
    \centering
    \begin{tabular}{c c c c}
        \hline
        \hline
        \textbf{Network load($\lambda$)} & $2$ & $6$ & $8$ \\
        \hline
        Shortest-path fixed one & 1.2 & 3.1 & 4.1 \\
        Shortest-path fixed two & 1.6 & 4.0 & 5.3 \\
        FIFO & 1.2 & 3.1 & 4.2 \\
        DNN medium & 1.8 & 4.7 & 6.1 \\
        DNN high & 1.9 & 5.0 & 6.5 \\
        Bayesian medium & 1.8 & 4.7 & 6.1 \\
        Bayesian high & 2.0 & 5.1 & 6.6 \\
        \hline
        \hline
    \end{tabular}
    \vspace{-0.5cm}
\end{table}

\begin{table}[ht]
	\caption{Gains in ``DNN medium" over ``Shortest-path fixed two".}
    \vspace{-0.25cm}
    \renewcommand{\arraystretch}{1.05}
    \addtolength{\tabcolsep}{10pt}
	\label{tab:gains}
	\centering 
	\begin{tabular}{c c c c}
        \hline
        \hline
        \textbf{Network load($\lambda$)} & $2$ & $6$ & $8$ \\
        \hline
        \textbf{Latencies \%} & $5.65$ & $6.30$ & $8.23$ \\
        \textbf{Success rates \%} & $11.94$ & $15.36$ & $14.31$ \\
        \hline
        \hline
	\end{tabular}        
\end{table}



\section{Conclusion and Future Work} \label{sec:conc}


In this paper, we propose hop-level purification-round estimators (DNN and Bayesian) integrated with a cost-based scheduler that jointly optimizes path selection and purification to achieve a target fidelity. The method is evaluated in latency, fidelity, success rate, and Bell pair utilization. Compared with fixed-round purification schemes, it offers adjustable fidelity levels, reduces latency via its cost function, and improves Bell pair utilization. Specifically, latency is reduced by up to 8\% and success rate increased by up to 14\%.


\section*{Acknowledgment}
This publication has emanated from research conducted with the financial support of Research Ireland under Grant number \text{13/RC/2077\_P2}. For the purpose of Open Access, the author has applied a CC BY public copyright licence to any Author Accepted Manuscript version arising from this submission.

\bibliographystyle{IEEEtran}

\bibliography{ref}

\end{document}